\documentclass{article}

\usepackage{amssymb}

\newtheorem {z}{Task}
\newtheorem {prop} {Proposition}

\title{An Entertaining Example of Using the Concepts of Context-Free Grammar and Pushdown Automation}

\author{Krasimir Yordzhev}
\date{}

\begin{document}
\maketitle

Faculty of Mathematics and Natural Sciences, South-West University

66 Ivan Mihailov Str, 2700 Blagoevgrad, Bulgaria
    
e-mail: yordzhev@swu.bg

\begin{abstract}
A formal-linguistic approach for solving an entertaining task is made in this paper. The well-known task of the Hanoi towers is discussed in relation to some concepts of discrete mathematics. A context-free grammar which generate an algorithm for solving this task is described.  A deterministic pushdown automation which in its work  imitates the work of monks in solving the task of the Hanoi towers is built.
\end{abstract}


\section{Introduction}

\begin{z} \label{KI_Hanoj} {\bf (The Task of the Hanoi Towers \cite{arsac})}  The Hanoi Towers are made up of three vertical columns.
On the first is hung a series of $N$ discs. The discs are all different, but ordered by size with the largest being on
the bottom and the smallest on top. The task is to move the discs from the first to the third column, using the second
column as an assistant. There are several conditions to completing this exercise: only one disc may be moved at one
time and while one disc is being moved, all other discs must be on one of the columns and also during this time
it is prohibited a larger disc to be on smaller.

\end{z}

The task of the Hanoi Towers is a classic example used to teach recursion in programming \cite{wirth,arsac}. In this paper, we will look at this Task from the standpoint of mathematical linguistics, i.e. as a part of the discipline of  ''discrete mathematics'' and ''discrete structures'' studies by students in Informatics and Computer courses at university \cite{dpd,dshtr,manev,syt}.
	
The algebraic properties of context-free grammars and languages are discussed in detail in \cite{ginzburg,laleman}. Several practical applications of formal grammars and languages and pushdown automata are discussed in detail in \cite{aho_ulman,hopcroft}.

\section{Context-free grammars and languages }

\hspace{0.25in}Let $V$ be a finite and non-empty set. The elements of this set we will call {\it letters}, and the whole set $V$ -
{\it Alphabet}.

{\it A word} over the alphabet $V$ we will call each finite string of letters by $V$.
The word that does not contain any letter is called {\it the empty word} that we will mark with $\varepsilon$. $V^*$ denotes the set of all words over $V$, including empty set.
 By{\it the length} of a word refers to the number of letters in it. The length of the word $\alpha$ will be expressed with $|\alpha |$.

Let $\alpha$ and $\beta$  be two words over the Alphabet $V$. By {\it concatenation (mu\-lti\-pli\-ca\-tion, adhesion)} $\alpha \beta$ of both words we will mean the word obtained by successive completion of the letters of
$\beta$ after the last letter of $\alpha$.

Let $V$ be alphabet. Each subset  $L$ of $V^*$ is called {\it
formal language} (or only {\it language})  over alphabet $V$.

By {\it generative grammar} (or only {\it grammar}) $\Gamma$ we will understand the four ordered tuples
$\Gamma =\langle V,W,S,P\rangle$, where $V$ is a finite set (Alphabet) from {\it terminal} symbols, $W$- set of {\it non - terminal} symbols,  $S$-{\it
start symbol} of the grammar, which is an element of $W$, and $P$ is set of ordered pairs $(\alpha ,\beta )$, where
$\alpha ,\beta \in (V\cup W)^*$, as in $\alpha $ there is at least one non - terminal symbol. 	
In a number of sources (see references in the end) and  an additional condition is placed while sets $W$ and $P$ to be finite.
For our needs this condition is not necessary, it is enough that these sets are countable. The elements of $P$ are called
{\it productions}. If $(\alpha ,\beta )\in P$, then it means $\alpha \to \beta$, as the symbol $"\to "$ does not belong to
$V\cup W$.

Let $\mu$ and $\nu$ be two words from $(V\cup W)^*$. We will say that
$\mu$ is {\it derived directly} from $\nu$ in the grammar
$\Gamma =\langle V,W,S,P\rangle$ and will write $\nu
\stackrel{\Gamma}{\vdash} \mu$ (or only $\nu \vdash \mu$ , if
$\Gamma$ is understandable), here exists words $\alpha_1 ,\alpha_2
\in (V\cup W)^*$ and production $\alpha \to \beta$ in $P$ so that $\nu
=\alpha_1 \alpha \alpha_2$ and $\mu =\alpha_1 \beta\alpha_2$.

If $\omega_0 ,\omega_1 ,\ldots ,\omega_n$ is a word over $V\cup W$,
for which $\omega_0 \stackrel{\Gamma}{\vdash} \omega_1
\stackrel{\Gamma}{\vdash}  \ldots
\stackrel{\Gamma}{\vdash} \omega_n$, we will say that number of words is
{\it the derivation} of $\omega_n$ from $\omega_1$ in $\Gamma$ which we denote with
$\omega_0 \stackrel{\Gamma}{\models} \omega_n$ or
only $\omega_0 \models \omega_n$, if $\Gamma$ is default.
The count $n$ of immediate derivations  $\omega_i
\stackrel{\Gamma}{\vdash} \omega_{i+1}$ will be called {\it the length of
derivation}.

The Set $L(\Gamma )=\{ \omega\in V^* \; |\;
S\stackrel{\Gamma}{\models} \omega \}$ is called {\it formal
language over $V$, phrase - structured grammar  $\Gamma$}.
The Grammars $\Gamma_1$ and $\Gamma_2$ are {\it
equivalent} if $L(\Gamma_1 )=L(\Gamma_2 )$.

A grammar $\Gamma =\langle V,W,S,P\rangle$ is
{\it context-free}, if all \`productions of it are from the type
$$A\to \omega ,\quad A\in V,\quad \omega \in (V\cup W)^* ,$$
where $V$ and $W$ are alphabets respectively with terminal and nonterminal symbols.

\begin{z} \label{KF-grammar}
For a given positive integer $N$  a context-free grammar $\Gamma_N$ should be built with terminal alphabet encoding
possible displacements and if $\omega \in L(\Gamma_N )$, then $\omega$  describes algorithm that so\-lve the Task  \ref{KI_Hanoj}.
Prove that for each positive integer $N$ language $L(\Gamma_N )$ is not empty, i.e for each positive integer $N$ there is an algorithm solves
the task of the Hanoi towers. \footnote{It is not necessary $L(\Gamma_N)$ to describe all solutions of Task \ref{KI_Hanoj}. Some of them may be ineffective, for instance if they involve the useless relocation of disk as soon as moving the same disk to another column.}.
\end{z}

Solution. Let's consider context-free grammar $\Gamma_N =\langle V,W,S,P\rangle$ where $V=\{ p_{ij} \; |\;
i,j\in \{1,2,3\} ,\; i\ne j\}$. The meaning of $p_{ij}$ is $"$Move top disk from the $i-th$ column of the $j-th$ column$"$.
In this way, if $\omega =\pi_1 \pi_2 \ldots \pi_k$, where $\pi_i \in V,$ $i=1,2,\ldots ,k$, then $\omega$ describes algorithm for moving consecutively
of $k$ discs in the three columns.

 $W=\{
h_{ij} (n)\; |\; i,j\in \{ 1,2,3\} ,\; i\ne j,\; n=1,2,\ldots ,N\} ,$
start symbol $S=h_{13} (N)$, $P$ consists of productions $h_{ij} (1)\to p_{ij}$ and
$h_{ij} (n) \to h_{ik} (n-1) p_{ij} h_{kj} (n-1)$ about $n=2,3,\ldots
,N$, where $i,j,k\in \{1,2,3\} ,$ $ i\ne j,$
$i\ne k,$ $ k\ne j$.
Apparently, so constructed grammar $\Gamma_N$ is context-free.

Let $\omega\in L(\Gamma_N )$, i.e. we assume that the derivation $h_{13} (N) \stackrel{\Gamma_N}{\models} \omega$ exist and let the length of derivation is equal of $s$. If $s=1$, then obviously  this is possible if and only if number of discs $N=1$ and we have direct derivation $h_{13} (1) \stackrel{\Gamma_N}{\vdash} p_{13}$ and in presence of single disc $p_{13}$ describes an algorithm for solving the Task \ref{KI_Hanoj}. Similarly  $h_{ij} (1) \stackrel{\Gamma_N}{\vdash} p_{ij}$ is derivation with length 1 with start symbol $h_{ij} (1)$ and $p_{ij}$ describes algorithm about moving single disc from column $i$ to column $j$, where $i,j\in \{ 1,2,3\}$ and  $i\ne j$.   Let $s>1$. We assume that if derivation exists with length less than  $s$ from type $h_{ij} (n) \stackrel{\Gamma_N}{\models} \alpha$, where $\alpha\in V^*$, $n\le N$ then $\alpha$ describes moving of $n$ discs from column $i$ on column  $j$, using column $k$ as assistant , according to constraints in Task \ref{KI_Hanoj}, where $i,j,k\in \{ 1,2,3\}$, $i\ne j,$ $j\ne k,$ $k\ne i$. When $s>1$ obviously derivation $h_{13} (N) \stackrel{\Gamma_N}{\models} \omega$  with length $s$ (if existing)it will have the type $h_{13} (N)\stackrel{\Gamma_N}{\vdash} h_{12} (N-1) p_{13}h_{23}(N-1) \stackrel{\Gamma_N}{\models} \omega$ then next derivations exist $h_{12} (N-1)\stackrel{\Gamma_N}{\models} \omega_1$ and $h_{23} (N-1) \stackrel{\Gamma_N}{\models} \omega_2$ with lengths less then $s$, where $\omega_1 ,\omega_2 \in V^*$ and $\omega =\omega_1 p_{13} \omega_2$. According to induction assumption $\omega_1$ describes algorithm for moving of $N-1$ discs from first to second column, using third one as assistant, and $\omega_2$ describes algorithm for moving of $N-1$ discs from second to third column, using first one as assistant and pay attention to constraints from Task \ref{KI_Hanoj}. Then $\omega =\omega_1 p_{13} \omega_2$ describes the following algorithm: first move while considering with constraints from Task  \ref{KI_Hanoj} the top number $t$ of discs from first column to second then move the largest bottom disk from the first column of the empty third and finally move $t$ number of discs from second column to third one. Therefore  $\omega =\omega_1 p_{13} \omega_2 $ (if existing ) describe the solution of the task of the Hanoi towers.

Let prove for each positive integer $N$ language $L(\Gamma_N )$ is not empty.  When $N=1$ the only production of $P$ which can be applied is $h_{13}\to p_{1 3}$  and therefore $L(\Gamma_1)=\{p_{13} \}$, i.e. $L(\Gamma_1 )$ is not empty language. Let's assume that for each positive integer  $t\le N$ the languages $L(\Gamma_t )$ are not empty and put $N=t+1$. Let consider the context-free grammar $\Gamma_t' =\langle V,W,h_{12} (t),P\rangle$ and $\Gamma_t'' =\langle V,W,h_{23} (t),P\rangle$. Apparently $\Gamma_t'$ and $\Gamma_t''$ work by analogy of  $\Gamma_t$ and according to the above proven if $\omega' \in L(\Gamma_t' )$, then $\omega' $ describes algorithm for moving of $t$ discs from first to second column, using third one as assistant according to constraints described in Task \ref{KI_Hanoj}, and if $\omega'' \in L(\Gamma_t'')$, then $\omega''$ describes the algorithm for moving of $t$ discs from second column to third one using first one as assistant. According to induction assumption $\omega'$ and $\omega''$ exist. Then in $\Gamma_{t+1}$ derivation exist $h_{13} (t+1) \stackrel{\Gamma_{t+1}}{\vdash}  h_{12} (t)p_{13} h_{23} (t) \stackrel{\Gamma_{t+1}}{\models} \omega' p_{13}\omega'' $, where $\omega' \in L(\Gamma_t')$ and $\omega'' \in L(\Gamma_t'')$. Therefore $\omega\in L(\Gamma_{t+1} )$, i.e. $L(\Gamma_{t+1} )$ is non empty language.

\hfill $\Box$

When $N=5$ next word is produced
$p_{13}$ $ p_{12}$ $ p_{32}$ $ p_{13}$ $ p_{21}$ $ p_{23}$ $
p_{13}$ $ p_{12}$ $ p_{32}$ $ p_{31}$ $ p_{21}$ $ p_{32}$ $
p_{13}$ $ p_{12}$ $ p_{32}$ $ p_{13}$ $ p_{21}$ $ p_{23}$ $
p_{13}$ $ p_{21}$ $ p_{32}$ $ p_{31}$ $ p_{21}$ $ p_{23}$ $
p_{13}$ $ p_{12}$ $ p_{32}$ $ p_{13}$ $ p_{21}$ $ p_{23}$ $
p_{13}$  We can verify the correctness of the algorithm using the example about the five consecutive cards.

It is easy to prove (eg. using induction on $N$) following
\begin{prop}
Let $N$ is a positive integer, and $\Gamma_N$ is defined as the solution of Task \ref{KF-grammar}  context-free grammar, then
$$\left| L(\Gamma_N ) \right| =1$$
and if $\omega\in L(\Gamma_N)$, then
$$\left| \omega \right| =2^N -1$$

\hfill $\Box$
\end{prop}

In other words, for each a positive integer $N$ grammar $\Gamma_N$ generates exactly one word that
describes an algorithm for solving the task of the Hanoi towers with exactly $2^N -1$ displacements
of the disks from one column to another.

\section{Pushdown automata}

\hspace{0.25in}By  {\it nondeterministic pushdown automation } one will understand each ordered septenary $$M=\langle
K,V,W,\delta ,q_0 ,z_0 ,F\rangle ,$$ where:

- $K$ is a finite set of states of automaton {\it (states of alphabet)};

- $V$ is a finite set of entry letters {\it (entry alphabet)};

- $W$ is a finite, non empty set of {\it stack symbols
 (stack alphabet)};

- $\delta : K\times (V\cup \{\varepsilon \} )\times W \to {\cal P}
(K\times W^* )$  \footnote{ As usual with ${\cal P} (A)$
is denoted the set of all subsets of the set $A$,
including the empty.}  is {\it a transition function};

- $q_0 \in K$ is {\it a start state} of automaton;

- $z_0 \in W$ is {\it a start} stack  symbol;

- $F\subseteq K$ is a set of {\it accepting }states.

{\it Configuration } of nondeterministic pushdown automaton $M$
that is  the ordered triple $\langle q,\alpha ,\gamma \rangle \in
K\times V^* \times W^*$.

 Let $\alpha =a_1 a_2 \ldots a_s \in V^* ,$ $\gamma =z_1 z_2
\ldots z_t \in W^*$. Then transition function $\delta$
defines transition configuration $\langle q,\alpha ,\gamma
\rangle $ to the next configuration in the following way:

$(i)$ for each pair $\langle p,\gamma ' \rangle \in\delta (q,
a_1 ,z_1 )$ the configuration  $\langle q,\alpha ,\gamma \rangle $
passing in the configuration  $\langle p,\alpha_1 ,\gamma_1 \rangle
$, where $\alpha_1 =a_2 a_3 \ldots a_s ,$ $\gamma_1 =\gamma ' z_2
z_3 \ldots z_t $, which we denote by  $\langle q,\alpha ,\gamma
\rangle \vdash \langle p,\alpha_1 ,\gamma_1 \rangle $.

$(ii)$ for each pair $\langle p,\gamma ' \rangle \in\delta (q,
\varepsilon ,z_1 )$ the configuration  $\langle q,\alpha ,\gamma
\rangle $ passing in the configuration   $\langle p,\alpha ,\gamma '
z_2 z_3 \ldots z_t \rangle$, which we denote by $\langle
q,\alpha ,\gamma \rangle \vdash \langle p,\alpha ,\gamma ' z_2 z_3
\ldots z_t \rangle$.

In the beginning if nondeterministic
pushdown automation  is given the word
$\alpha =a_1 a_2 \ldots a_s$, then according to the start
configuration $\langle q_0 ,\alpha ,z_0 \rangle $
the following possible configurations are obtained by using a function of transitions
$\delta$. For each new configuration using  $\delta$  all
possible next configurations are obtained and so on.

 Nondeterministic pushdown automation $M$ {\it recognizes the word  $\alpha =a_1 a_2 \ldots a_s$ by accepting st\-ate,}
  if its work at the beginning of given word $\alpha$,
it reaches a con\-fi\-gu\-ra\-tion of type  $\langle q ,\varepsilon
,\gamma \rangle $, for each $\gamma \in W^*$, when $q\in F$.

 Nondeterministic pushdown automation $M$ {\it recognizes the word
  $\alpha =a_1 a_2 \ldots a_s$ by empty stack,} if its work 	
at the beginning of given word $\alpha$, reaches a con\-fi\-gu\-ra\-tion of type $\langle q ,\varepsilon
,\varepsilon \rangle $.

 Pushdown automation $M=\langle K,V,W,\delta ,q_0 ,z_0
,F\rangle$ is called {\it deterministic}, if for each $q\in
K$ and $z\in W$ or

(1) $\delta (q,a,z)$ contains no more than one element for each  $a\in V$ and $\delta (q,\varepsilon ,z)=\emptyset$\\
or

(2) $\delta (q,a,z)=\emptyset$ for each  $a\in V$ and $\delta
(q,\varepsilon ,z)$ contains no more than one element.

 Language which is recognized by some deterministic pushdown automation is called {\it
deterministic} language.
As it is known the relationship between context-free languages and pushdown automation
is given by next statements:

For each context-free language $L$ nondeterministic pushdown automation $M$ exists, such that $L$ is recognized by $M$ by accepting state.
Language $L$ is recognized by nondeterministic pushdown automation through an empty stack then only if $L$ is recognized by nondeterministic pushdown automation by accepting state.
If $L$ is a language which is recognized by a nondeterministic pushdown automation, then $L$ is context-free language.

\begin{z} \label{KI_Hanoj_MP}  For each positive integer  $N\ge 2$ deterministic pushdown automation $M_N$ should be built,
which in its work to imitate the work of monks in solving the task of the Hanoi towers. (see Task  \ref{KI_Hanoj}).
\end{z}

Solution. The requested pushdown automation is the following: $M=\langle K,\emptyset ,W,\delta ,q_0 ,z_0 ,\emptyset \rangle$, where $K=\{ q_0  \} ,$  $W=\{ p_{ij} \; |\; i\ne j,\; i,j=1,2,3\} \cup \{ h_{ij} (n)\; |\; i\ne j,\; i,j=1,2,3,\; n=1,2,\ldots ,N-1\} \cup \{ z_0 \}$,  and $\emptyset $ as usually the empty set is known . Let $i,j,k\in \{ 1,2,3\}$, $i\ne j,$ $j\ne k,$ $k\ne i$. Then the transition function $\delta$ is defined in following way:
\begin{equation}\label{1}
\delta (q_0 ,\varepsilon , z_0 )=\langle q_0 ,h_{12} (N-1)p_{13} h_{23} (N-1) \rangle
\end{equation}
\begin{equation}\label{2}
\delta (q_0 ,\varepsilon ,h_{ij} (1) )=\langle q_0 ,p_{ij} \rangle
\end{equation}
\begin{equation}\label{3}
\delta (q_0 ,\varepsilon ,h_{ij} (n) )=\langle q_0 ,h_{ik} (n-1)p_{ij} h_{kj} (n-1)\rangle \quad n=2,3,\ldots ,N
\end{equation}
\begin{equation}\label{4}
\delta (q_0 ,\varepsilon ,p_{ij} )=\langle q_0 ,\varepsilon \rangle .
\end{equation}

Immediately after inclusion of $M_N$, before being submitted as any input signal, the automation replaces the start stack symbol $z_0$  with word $h_{12} (N-1) p_{13} h_{23}(N-1)$ according to (\ref{1}) and after a number of actions depending on the current stack symbol (\ref{2}), (\ref{3}), or (\ref{4}). Moreover, we assume that automation  $M_N$  is designed so that after reading of stack symbol of the type $p_{ij}$ $i,j\in \{1,2,3\}$, $i\ne j$,  simultaneously according the action(\ref{4}) another action carried out namely the removal of top
disk $i$-th column on $j$-th. Transient function is defined so that after a finite number of beats the stack is empty and stops. This is because if the current stack symbol of the kind $p_{ij}$, then $M_N$  deletes it, and if the current stack symbol is of the type $h_{ij}(n)$, then at the next beat of the parameter $n$ decrease with one unit, if $n>1$, or $h_{ij}(n)$ passes into the symbol   $p_{ij}$ at $n=1$, then that symbol is deleted.

We will prove that $1\le s<N$ the stack $M_N$ of configuration $\langle q_0 ,\varepsilon ,h_{ij} (s)\gamma \rangle$ where $i,j\in \{ 1,2,3\} $, $i\ne j$, $\gamma\in W^*$ as a result of their work reaches a configuration $\langle q_0 ,\varepsilon ,\gamma \rangle$ and
when that transfer according to the restrictions of Task \ref{KI_Hanoj} $s$ discs from column $i$ to column $j$. When $s=1$ we have $\langle q_0 ,\varepsilon ,h_{ij} (1) \gamma \rangle \vdash \langle q_0 ,\varepsilon ,p_{ij} \gamma \rangle \vdash \langle q_0 ,\varepsilon ,\gamma \rangle$, i.e. the assertion is met. Assume that the assertion is fulfilled for any  $t$, such that  $1\le t<s<N$ and let $t=s$. Then in $i,j,k\in \{ 1,2,3\}$, $i\ne j$, $j\ne k$, $k\ne i$ we have $\langle q_0,\varepsilon , h_{ij} (t) \gamma \rangle \vdash \langle q_0 ,\varepsilon , h_{ik} (t-1)p_{ij} h_{kj} (t-1)\gamma \rangle$. According to induction assumption  $M_N$ reaches the configuration  $\langle q_0 ,\varepsilon ,p_{ij} h_{kj} \gamma \rangle$ moving $t-1$ discs from $i$-th column of $k$-th column then passed in a configuration $\langle q_0 ,\varepsilon , h_{kj} (t-1)\gamma \rangle$ moving next disc from column $i$ to column $j$ and again according to the induction moves $t-1$ discs (as obviously all are smaller
size) on this disk column  $j$ taken from column $k$. Therefore the assertion is true for any $s=1,2,\ldots ,N-1$.

According the assertion that has just proved we have:\\
$\langle q_0 ,\varepsilon , z_0 \rangle \vdash \langle q_0 ,\varepsilon ,h_{12} (N-1) p_{13}h_{23} (N-1)  \rangle \vdash \ldots \vdash \langle q_0 ,\varepsilon , p_{13} h_{23} (N-1) \rangle \vdash \langle q_0 ,\varepsilon , h_{23} (N-1) \rangle \vdash \ldots \vdash \langle q_0 ,\varepsilon ,\varepsilon  \rangle $, 	
while the stack moves at the upper  $N-1$ discs from first to second column, then move biggest on the bottom  from the first to third column and finally moving discs  ($N-1$ numbers from the second to third column while observed the restrictions described in the Task \ref{KI_Hanoj}. Therefore pushdown automation $M_N$ solve the task of the Hanoi towers.

\hfill $\Box$

\bibliographystyle{plain}
\bibliography{kybib}

\end{document}